# Mapping quantum random walks onto a Markov chain by mapping a unitary transformation to a higher dimension of an irreducible matrix

Arie Bar-Haim

Davidson Institute of Science Education, Israel

Here, a new two-dimensional process, discrete in time and space, that yields the results of both a random walk and a quantum random walk, is introduced. This model describes the population distribution of four coin states $|1\rangle, -|1\rangle, |0\rangle, -|0\rangle$ in space without interference, instead of two coin states $|1\rangle, |0\rangle$. For the case of no boundary conditions, the model is similar to a Markov chain with a stochastic matrix, i.e., it conserves the population distribution of the four coin states, and by using a proper transformation, yield probability distributions of the two quantum states $|1\rangle, |0\rangle$ in space, similar to a unitary operator. Numerical results for a quantum random walk on infinite and finite lines are introduced.

Various problems of quantum random walks have been investigated by many groups. For example, Aharonov et al. [1] explored quantum random walks, while Ambainis et al. [2–4] examined quantum walks on graphs. Bach et al. [5] investigated one-dimensional quantum walks with absorbing boundary conditions. Dür et al. [6] discussed quantum random walks in optical lattices. Moreover, Konno et al. [7] examined absorption problems and the eigenvalues of two-state quantum walks [8]. Mackay et al. [9] explored quantum walks in higher dimensions and Bartlet et al. [10] examined quantum topology identification in addition to various other problems [11,12].

Several studies have discussed the differences between random walks and quantum random walks, such as those conducted by Childs et al. [13] and Motes et al. [14] . The differences are manifested in various distribution functions that influence the moments of the dynamics. For example, the probability distribution function of a symmetric random walk starting at the origin behaves like a Gaussian distribution around the origin, i.e., there is a high probability of the walker being found at the origin, while in the case of a quantum random walk, there is a low probability of being found at the origin, mainly because of interference [1,7].

In the present study, a new two-dimensional model, discrete in time and space, that yields the results of both a random walk and quantum random walk, is introduced. This model describes the population distribution of four coin states $|1>, -|1>, |0>, -|0>$ in space without interference.. Using a proper transformation (introduced below) on the four coin state distributions yields an amplitude distributions for the two quantum states, $|1>, |0>$, which is similar to a unitary operator. The model shows that the asymptotic behavior of the distribution of each of the four coin states behaves like a Gaussian distribution. The model also enables us to extend it to different boundary conditions, such as a reflecting point or a trap, as well as to obtain the distribution of each of the four coin states separately, thereby gaining a better understanding of the entire system.

# Equivalence between a Hadamard operator and a Markov chain

The discrete-time quantum random walk is defined by two operators: the coin flip operator and the shift operator, where the Hilbert space that governs the walk is a tensor product of $\mathcal{HT} = \mathcal{HC} \otimes \mathcal{HS}$, so that $\mathcal{HC}$ is a Hilbert space that can be defined by the two canonical bases $|0>$ and $|1>$ as follows:

$$|0>= \begin{pmatrix}1\\0\end{pmatrix} \quad |1>= \begin{pmatrix}0\\1\end{pmatrix}, \tag{1}$$

and $\mathcal{HS}$ is a Hilbert space defined by an infinite canonical base
$|k>= (0,0,0,\ldots 1,0,0)^T$, where 1 stands for the kth position in space.

The amplitude of a particle at the kth location at any time step $n$, defined by a 2D vector, is as follows:

$$|\psi_k(n)>= \alpha|0>_c + \beta|1>_c \tag{2}$$

where the subscript c denotes that these states belong to the $\mathcal{HC}$ space.

The probability that the particle is at location $k$ at time step $n$ is given by the square of the modulus of $|\psi_k(n)>$, namely, $||\psi_k(n)>|^2$

The Hadamard operator is defined by the following unitary operator:

$$H = \frac{1}{\sqrt{2}} H_d, \tag{3}$$

where $H_d$ is the Hadamard matrix:

$$H_d = \begin{bmatrix}1 & 1\\1 & -1\end{bmatrix}, \tag{4}$$

In order to build our model, the $H$ operator was applied to the following coin states:
$|0>_c, |1>_c, -|1>_c, -|0>_c$, which yields

$$H|0>_c = \frac{\sqrt{2}}{2}(|0>_c + |1>_c), \tag{5}$$

$$H|1>_c = \frac{\sqrt{2}}{2}(|0>_c - |1>_c), \tag{6}$$

$$H(-|1>_c) = \frac{\sqrt{2}}{2}(-|0>_c + |1>_c), \tag{7}$$

$$H(-|0>_c) = \frac{\sqrt{2}}{2}(-|0>_c - |1>_c), \tag{8}$$

Using these results, the Hadamard operator can be described by the following Markov chain with two reflecting points and transition probabilities of $p = q = 0.5$, as shown in Figure 1.

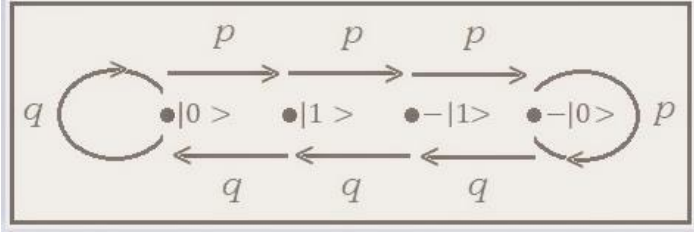

Figure 1. A four-site Markov chain that represents the transitions between the coin states $|0>_c, |1>_c, -|1>_c, -|0>_c$.

The analogy between a Hadamard operator and the model depicted in Figure 1 is explained as follows:

When starting at coin state $|0>_c$, the walker can stay at $|0>_c$ or jump to state $|1>_c$ (Eq. (5)).

When starting at coin state $|1>_c$, the walker can jump backward to state $|0>_c$ or jump to state $-|1>_c$ (Eq. (6)).

When starting at coin state $-|1>_c$, the walker can jump backward to state $|1>_c$ or jump to state $-|0>_c$ (Eq. (7)).

When starting at coin state $-|0>_c$, the walker can stay at that state or jump to state $-|1>_c$ (Eq. (8)).

Stated formally: the transition probabilities matrix that represents the Markov chain depicted in Figure 1 is as follows:

$$A = \begin{bmatrix} 0.5 & 0.5 & 0 & 0 \\ 0.5 & 0 & 0.5 & 0 \\ 0 & 0.5 & 0 & 0.5 \\ 0 & 0 & 0.5 & 0.5 \end{bmatrix}, \qquad (9)$$

In order to present the mathematical relation between the transition matrix $A$ and the Hadamard operator, the interference matrix $B$ is defined as follows:

$$B = \begin{bmatrix} 1 & 0 & 0 & -1 \\ 0 & 1 & -1 & 0 \end{bmatrix}, \qquad (10)$$

Several properties of the interference matrix will be utilized, including the following:

$$BB^T = 2\begin{bmatrix} 1 & 0 \\ 0 & 1 \end{bmatrix} = 2I_2, \tag{11}$$

and

$$B^T B = \begin{bmatrix} 1 & 0 & 0 & -1 \\ 0 & 1 & -1 & 0 \\ 0 & -1 & 1 & 0 \\ -1 & 0 & 0 & 1 \end{bmatrix} = I_4 - J_4, \tag{12}$$

where $I_4$ is the identity matrix, $J_4$ is a reversal matrix, and the subscript denotes its dimension. Using the interference matrix $B$, the Hadamard matrix can be presented as

$$H = \frac{1}{\sqrt{2}} BAB^T. \tag{13}$$

which can be shown explicitly by multiplication as follows:

$$\frac{1}{\sqrt{2}}\begin{bmatrix} 1 & 0 & 0 & -1 \\ 0 & 1 & -1 & 0 \end{bmatrix} \begin{bmatrix} 0.5 & 0.5 & 0 & 0 \\ 0.5 & 0 & 0.5 & 0 \\ 0 & 0.5 & 0 & 0.5 \\ 0 & 0 & 0.5 & 0.5 \end{bmatrix} \begin{bmatrix} 1 & 0 \\ 0 & 1 \\ 0 & -1 \\ -1 & 0 \end{bmatrix} = \frac{1}{\sqrt{2}}\begin{bmatrix} 1 & 1 \\ 1 & -1 \end{bmatrix}. \tag{14}$$

In other words, the unitary Hadamard operator is mapped onto a higher dimension of a symmetric Markov chain (Note that $\text{Det}(H) \neq 0$, while $\text{Det}(A) = 0$).

## The power of $A^n$

The mathematical relation between $A^n$ and $H^n$, where $n$ is the step number, is explained in the following equations:

Based on Eq. (13), $H^2 B$ can be written as

$$H^2 B = \left(\frac{1}{\sqrt{2}}\right)^2 (BAB^T)(BAB^T)B, \tag{15}$$

Since $B^T B$ commutes with the transition matrix $A$ (it can be shown explicitly by matrix multiplication: $B^T BA - AB^T B = 0$ furthermore note that: $B^T B = I_4 - J_4$ and both the identity matrix, $I_4$, and the reversal matrix, $J_4$, commute with A). Therefore Eq. (15) can be rearranged to

$$H^2 B = \left(\frac{1}{\sqrt{2}}\right)^2 B(B^T B)^2 A^2, \tag{16}$$

and similar to Eq. (16), the following relation can be written in general for any positive integer $n$:

$$H^n B = \left(\frac{1}{\sqrt{2}}\right)^n B(B^T B)^n A^n, \tag{17}$$

Note that $B^T B/2 = \frac{1}{2}(I_4 - J_4)$ from Eq. (12), and $(B^T B/2)^2$ satisfies

$$(B^T B/2)^2 = \frac{1}{4}(I_4 I_4 - I_4 J_4 - J_4 I_4 + J_4 J_4) = \frac{1}{4}(I_4 - 2J_4 + I_4) = \frac{1}{2}(I_4 - J_4), \tag{18}$$

which means that $B^T B/2$ is an idempotent matrix (M is idempotent if $M^2 = M$), and in general, $(B^T B/2)^n = B^T B/2$. Thus this yields:

$$(B^T B)^n = 2^{n-1} B^T B \tag{19}$$

Substituting Eq. (19) into Eq. (17) yields:

$$H^n B = \left(\frac{1}{\sqrt{2}}\right)^n 2^{n-1} B B^T B A^n, \tag{20}$$

and using the property of Eq. 11, $BB^T = 2I$, yields the desired relation:

$$H^n B = B\sqrt{2}^n A^n. \tag{21}$$

Consider the following arbitrary initial condition of the Markov chain:

$$P(0) = \begin{pmatrix} \alpha \\ \beta \\ \gamma \\ \delta \end{pmatrix}, \tag{22}$$

where $P(0)$ describes the four coin states: $|0>_c, |1>_c, -|1>_c, -|0>_c$ at step $n = 0$.

This initial condition corresponds to

$$|\psi(0)> = \alpha|0>_c + \beta|1>_c - \gamma|1>_c - \delta|0>_c, \tag{23}$$

Applying Eq. (21) to the initial condition yields

$$H^n B P(0) = B\sqrt{2}^n A^n P(0), \tag{24}$$

Rearranging the left side yields

$$H^n |\psi(0)> = B\sqrt{2}^n A^n P(0), \tag{25}$$

and in general yields

$$|\psi(n+1)> = H|\psi(n)> = B\sqrt{2}^{n+1} AP(n). \tag{26}$$

Note that:

(a) The interference matrix operates only at the end of the process.

(b) The solution of the Markov chain conserves the distribution of the four coin states:

$$[1,1,1,1]P(n) = [1,1,1,1]P(0). \tag{27}$$

(c) By using the interference matrix, the squared norm of the quantum state $|\psi(n)>$ is preserved as a unitary operator:

$$2^n \|BP(n)\|^2 = \|\psi(n)\|^2 = \|\psi(0)\|^2 \tag{28}$$

## A random walk and a quantum random walk model

The process of a quantum random walk includes a shift operator [1], which is a unitary operator that acts on a different Hilbert space, $HS$. The two processes, Hadamard and the shift operators, can be described by model depicted in Figure 2, where the horizontal direction (x-axis) describes the movements due to the shift operator, and the vertical direction (y-axis) describes the movements due to the Hadamard operator, which are described by the transition matrix $A$.

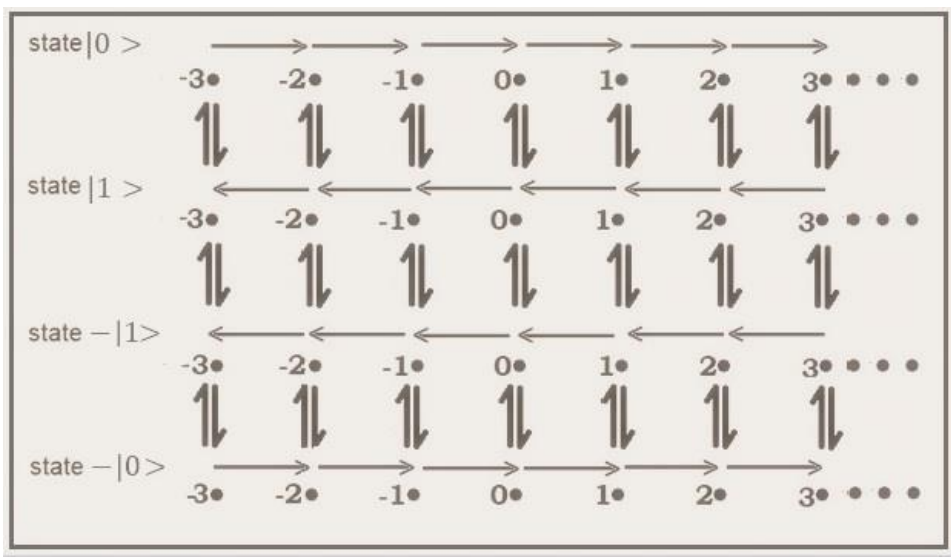

Figure 2. A two-step Markovian model. The horizontal direction (x-axis) describes the movements due to the shift operator, and the vertical direction (y-axis) describes the movements due to the transition matrix $A$.

The main points of the four-state Markov model are as follows:
1. The shift operator is responsible for the movement in the horizontal direction (x-axis), described by a birth-and-death process.
2. The Hadamard operator, presented by transition matrix $A$, is responsible for movements in the vertical direction (y-axis).
3. The movements in the horizontal direction and vertical direction occur one after the other, rather than simultaneously.

4. The same state can be populated without interference. Thus, it is possible to propagate $|0>_c, -|0>_c$ at the same kth site.
5. Interference occurs at the end of the dynamics.
6. In order to obtain the amplitude of the quantum state after $n$ steps, the distribution of states is multiplied by a factor of $\sqrt{2}^n$ (this factor will be explained in the following section)

Stated formally: the Hadmarad walk can be formulated as follows:

$$y = I_d \otimes H \tag{29}$$

where the subscript $d$ denotes the size of the unit matrix $(dXd)$ and the shift operator, x, can be presented as:

$$x = Right \otimes Zero + Left \otimes One \tag{30}$$

where:

$$Zero = \begin{bmatrix} 1 & 0 \\ 0 & 0 \end{bmatrix}, One = \begin{bmatrix} 0 & 0 \\ 0 & 1 \end{bmatrix} \tag{31}$$

The *Right* and the *Left* matrices are $d \times d$ zero matrices, except those entries that appear above the main diagonal and those that appear beneath the main diagonal, respectively:

$$Right\ (j, j+1) = 1, Left(j+1, j) = 1 \tag{32}$$

for integer $j$ between 1 to $d-1$ (inclusive)

The dynamics of the Hadmarad walk can be written as $u = xy$, therefore multiplying Eq. (30) by Eq. (29) yields:

$$u = (Right \otimes Zero + Left \otimes One)\ I_n \otimes H = Right \otimes ZeroH + Left \otimes OneH \tag{33}$$

The dynamic of the Markov chain depicted in Figure 2 is formulated similarly. The y-axis operator can be written as:

$$Y = I_d \otimes A \tag{34}$$

and the x-axis operators can be written as:

$$X = Right \otimes Zerostate + Left \otimes OneState \tag{35}$$

where the matrices Zerostate and Onestate replace the matrices Zero and One in Eq. (30)

$$Zerostate = \begin{bmatrix} 1 & 0 & 0 & 0 \\ 0 & 0 & 0 & 0 \\ 0 & 0 & 0 & 0 \\ 0 & 0 & 0 & 1 \end{bmatrix}, Onestate = \begin{bmatrix} 0 & 0 & 0 & 0 \\ 0 & 1 & 0 & 0 \\ 0 & 0 & 1 & 0 \\ 0 & 0 & 0 & 0 \end{bmatrix} \tag{36}$$

The dynamic of a complete step of the Markov chain depicted in Figure 2 is:

$$U = XY = Right \otimes ZerostateA + Left \otimes OneStateA \qquad (37)$$

The equivalence between the two systems is proven step-by-step beginning by substituting $H = \frac{1}{\sqrt{2}} BAB^T$ into Eq. (29) which obtains:

$$y = I_d \otimes H = \frac{1}{\sqrt{2}}(I_d \otimes BAB^T) \qquad (38)$$

The shift operator $x$, described by Eq. (30) can be written in terms of the matrices $ZerostateA, OneState$ as follows:

$$x = Right \otimes Zero + Left \otimes One = \qquad (39)$$
$$\frac{1}{2}(Right \otimes BZerostateB^T + Left \otimes BOneStateB^T)$$

Since $BZerostateB^T = 2Zero$ and $BOneStateB^T = 2One$ (which can be shown explicitly by multiplication), the entire dynamics of the Hadamard walk can be formulated as:

$$u = xy = \frac{1}{2}(right \otimes BZerostateB^T + left \otimes BOneStateB^T) \frac{1}{\sqrt{2}}(I_d \otimes BAB^T) \qquad (40)$$

Using the Kronecker product property of $(a \otimes b)(c \otimes d) = ac \otimes bd$ on Eq. (40) yields

$$u = xy = \frac{1}{2\sqrt{2}}(right \otimes BZerostateB^T BAB^T + left \otimes BOneStateB^T B AB^T) \qquad (41)$$

(The dimension of the right and left matrices are the same as the identity matrix)

Continuing with the multiplication of both sides by $(I_d \otimes B)$ yields:

$$u(I_d \otimes B) = \frac{1}{2\sqrt{2}}(right \otimes BZerostateB^T BAB^T B + left \otimes BOneStateB^T B AB^T B) \qquad (42)$$

Note that the Kronecker product property is used for the right side of Eq. (42)

since $[Zerostate, B^T B] = 0$ and $[OneState, B^T B] = 0$ [15] and also because $[A, B^T B] = 0$, as was discussed before. Consequently, Eq. (42) can be rearranged as follows:

$$u(I_d \otimes B) = \frac{1}{2\sqrt{2}}(right \otimes BB^T BB^T BZerostateA + left \otimes BB^T BB^T BOneState\,A) \qquad (43)$$

Now, using the properties of $BB^T = 2I_2$ yields

$$u(I_d \otimes B) = \sqrt{2}(right \otimes BZerostateA + left \otimes BOneState\,A) \qquad (44)$$

and rearranging the last equation using the product property of Kronecker yields:

$$u(I_d \otimes B) = \sqrt{2}(I_d \otimes B)(right \otimes ZerostateA + left \otimes OneState\,A) \qquad (45)$$

Substituting The value of $U$ using Eq.(37) into Eq. (45) yields the relation ship between the Hadamard walk with the system depicted in Figure 2:

$$\underline{u(I_d \otimes B) = \sqrt{2}(I_d \otimes B)U} \quad (46)$$

In the following section, this relation is extended to obtain a general relationship as a function of step number $n$.

Multiplying both sides of Eq. (46) by $I_d \otimes B^T$ yields:

$$u = \frac{1}{\sqrt{2}}(I_d \otimes B)U(I_d \otimes B^T) \quad (47)$$

As a consequence, $u^2$ can be formulated explicitly as:

$$u^2 = \frac{1}{\sqrt{2}}(I_d \otimes B)U(I_n \otimes B^T)\frac{1}{\sqrt{2}}(I_d \otimes B)U(I_d \otimes B^T) \quad (48)$$

Next, multiplying both sides by $I \otimes B$ yields:

$$u^2(I_d \otimes B) = \frac{1}{\sqrt{2}}(I_d \otimes B)U(I_d \otimes B^T)\frac{1}{\sqrt{2}}(I_d \otimes B)U(I_d \otimes B^T)(I_d \otimes B) \quad (49)$$

and since $U$ commutes as: $U(I_d \otimes B^T)(I_d \otimes B) = (I_d \otimes B^T)(I_d \otimes B)U$ [16]

and $(I_d \otimes B)(I_d \otimes B^T) = I_d \otimes BB^T = 2(I_d \otimes I_2)$, then:

$$u^2(I_d \otimes B) = \left(\sqrt{2}\right)^2 (I_d \otimes B)U^2 \quad (50)$$

Therefore in general, the relationship between the two systems can be written as:

$$u^n(I_d \otimes B) = \left(\sqrt{2}\right)^n (I_d \otimes B)U^n \quad (51)$$

Finally, the following equation defines the equivalence between Hadamard Walk and the Markov chain depicted in Figure(2):

$$\underline{u^n(I_d \otimes B)P(0) = \left(\sqrt{2}\right)^n (I_d \otimes B)U^n P(0)} \quad (52)$$

The initial condition $P(0)$ is a d-dimensional column vector, where $d = 4m$ and m is the number of sites. $P(0)$ has entries described by $[p_1(0), p_2(0), p_3(n0), p_4(n0), p_5(0), p_6(0) ...]$. This vector unfolds in such a way that the first four entries describe the four coin states of the first site, the next four entries describe the four coin states of the second site, and so on. Note that $(I_d \otimes B)P(0)$ reduced the column vector to a $2m$ dimensional column vector, as expected for the system on the left side. In such a case, the first two entries describe the amplitude of $|0>_c, |1>_c$ quantum states of the first site, and the next two entries describe the quantum states of the second site, and so on. Note that Eq. (52) can be written as

$$\underline{u^n(I_d \otimes B)P(0) = \left(\sqrt{2}\right)^n (I_d \otimes B)P(n)} \quad (53)$$

where $P(n)$ is the solution of the system depicted in Figure 2, specifically: $P(n) = U^n P(0)$

Unfolding the vector $P(n)$ in the following way yields :

$$P_{|0>}(n) = [\, p_1(n), p_{1+4m}(n), p_{1+8m}(n), p_{1+12m}(n) \ldots p_{d-3}(n)\,] \tag{54}$$

$$P_{|1>}(n) = [\, p_2(n), p_{2+4m}(n), p_{2+8m}(n), p_{2+12m}(n) \ldots p_{d-2}(n)\,] \tag{55}$$

$$P_{-|1>}(n) = [\, p_3(n), p_{2+4m}(n), p_{3+8m}(n), p_{3+12m}(n) \ldots p_{d-1}(n)\,] \tag{56}$$

$$P_{-|0>}(n) = [\, p_4(n), p_{4+4m}(n), p_{4+8m}(n), p_{4+12m}(n) \ldots p_d(n)\,] \tag{57}$$

where $P_{|0>}(n), P_{|1>}(n), (n), P_{-|0>}(n)$ describes the amplitude distribution of the states $|0>, |1>, -|1>, -|0>$, respectively. $m$ is the site number and $p_j(n)$ is the jth entry of the vector $P(n)$ (The first entry of each of the vectors described by Eq. (54-57) presents the amplitude of the first site, the second entry presents the amplitude of the second site, and so on).

The probability distribution of the quantum states $|0>_c, |1>_c$ after $n$ steps are respectively:

$$\left|\psi_{|0>}(n)\right|^2 = 2^n \left|P_{|0>}(n) - P_{-|0>}(n)\right|^2 \tag{58}$$

$$\left|\psi_{|1>}(n)\right|^2 = 2^n \left|P_{|1>}(n) - P_{-|1>}(n)\right|^2 \tag{59}$$

Note that from the stochastic property :

$$\sum P(n) = \sum P(0) \tag{60}$$

and from the unitary property, the vector length is preserved. Thus, the squared norm that describes the sum of the probabilities distribution is:

$$2^n \left\|P_{|0>}(n) - P_{-|0>}(n)\right\|^2 + 2^n \left\|P_{|1>}(n) - P_{-|1>}(n)\right\|^2 \tag{61}$$
$$= \left\|P_{|0>}(n) - P_{-|0>}(n)\right\|^2 + \left\|P_{|1>}(0) - P_{-|1>}(0)\right\|^2$$

The following figures describe new aspects of the system obtained from the Markov process. Figures 3 and 4 present a numerical result after 100 steps ($n = 100$), starting at the origin in state $|0>_c$. Although it is a simple case, the numerical results provide us with additional information about the distributions of each of the four coin states and show that the distribution behaves like a Gaussian one, and only when interference occurs are the expected results of the quantum walk obtained.

The three graphs in the first row describe:

(a) The Markov distribution of state |0> in space after 100 steps, described by $P_{|0>}(n)$

(b) The Markov distribution of state |0> in space after 100 steps described by $P_{-|0>}(n)$

(c) The probability distribution of the quantum state |0> being at the kth site described by $|\psi_{|0>}(n)|^2$

Similarly, the three graphs in the second row describe:

(d) The distribution of state |1> in space after 100 steps described by $P_{|1>}(n)$

(e) The distribution of state –|1> in space after 100 steps described by $P_{-|1>}(n)$

(f) The probability distribution of the quantum state |1> being at the kth site described by $|\psi_{|1>}(n)|^2$

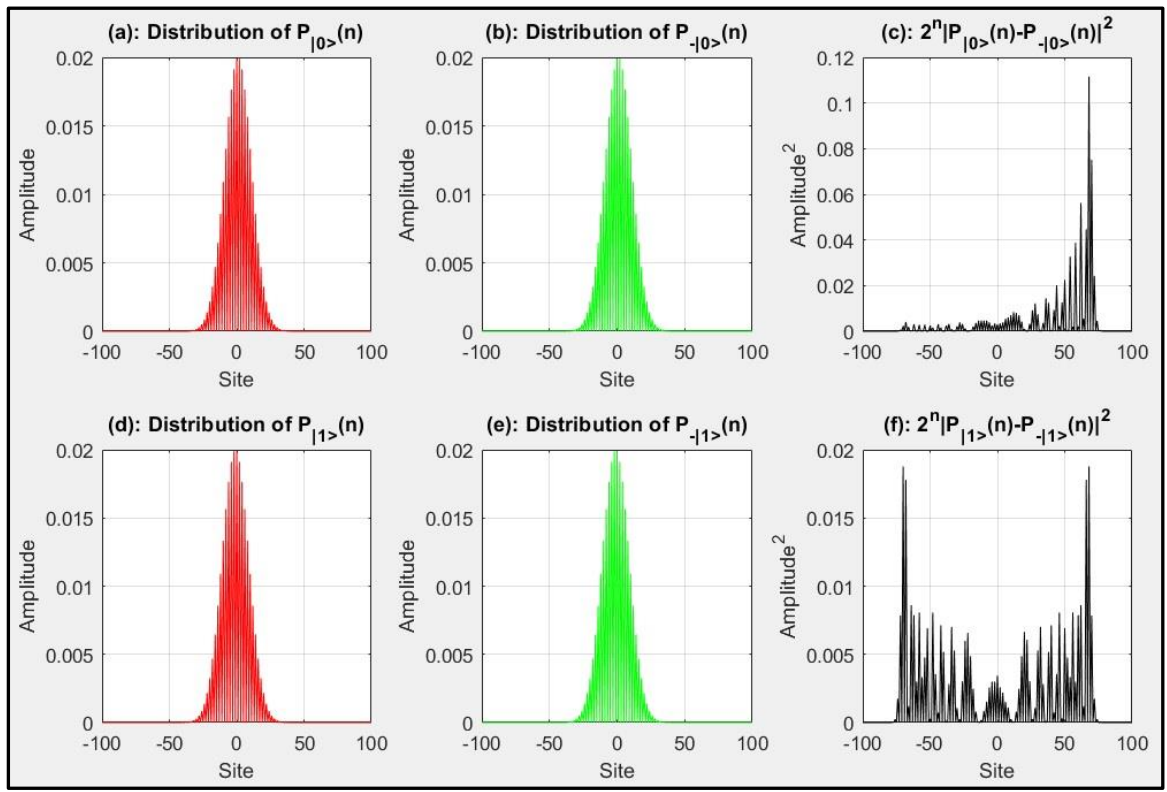

Figure 3. Graphs (a), (b), (d), and (e) present the Markov chain distribution of the four coin states $|0>_c, -|0>_c, -|1>_c, -|1>_c$, respectively. Graphs (c) and (f) present the probability distribution of the quantum states $|0>_c, |1>_c$.

Figure 3 shows that prior to interference, the distributions of the four states behave as Gaussian distributions. Adding the distributions depicted in Figure 3a,b,c,d yields the probability of the random walk being at the kth site (described as a black line in Figure 4), and adding the probability distributions of Figure 3c,f yields the well-known results of the probability of the quantum random walk being at the kth site (depicted as the blue line in Figure 4).

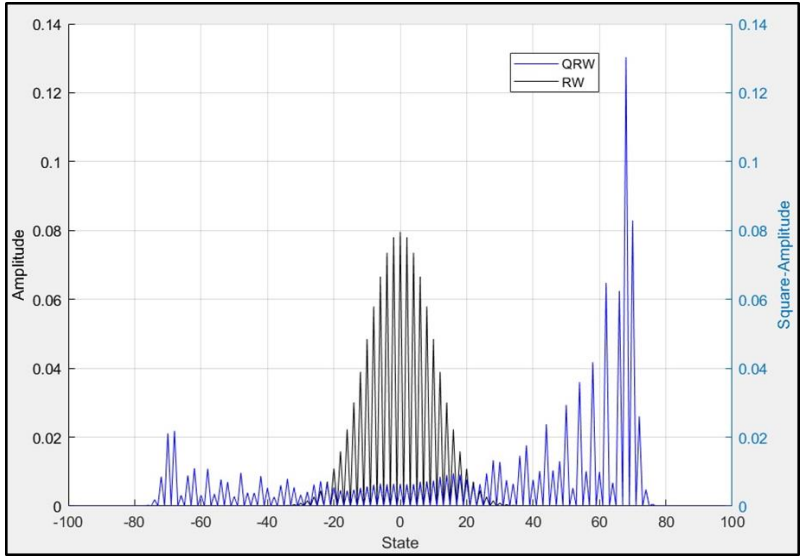

Figure 4. A random walk vs. a quantum random walk after 100 steps, starting at the origin.

The next figure presents a numerical result after 50 steps, initialized at the origin with $(|0>_c + i|1>_c)/\sqrt{2}$. In this case, the Markov process propagates two independent walkers: The first one is obtained by the *real part of* $(P(0))$ and the second one by the imaginary part of $(P(0))$. Both walkers (the real and imaginary part of $P(0)$) propagate at the same time, and thus each of the four distributions, $P_{|0>}(n), P_{|1>}(n), (n), P_{-|0>}(n)$, has a complex amplitude. The phase of each element of vector $P(n)$ can be defined as follows:

$$\theta_j(n) = tan^{-1}(image(p_j(n))/real(p_j(n))) \qquad (62)$$

(In the case of $real(p_j(n)) = 0$ then $\theta_j(n) = pi$ or $\theta_j(n) = -pi$, depending on the sign of the numerator; When the numerator and denominator are both zero, the phase is defined as zero), and the vector phase of $P_{|0>}(n)$ can be written as:

$$\Theta_{|0>}(n) = [\theta_1(n), \theta_{1+4m}(n), \theta_{1+8m}(n), \theta_{1+12m}(n) \dots \quad p\theta_{d-3}(n)] \qquad (63)$$

$$\Theta_{-|0>}(n) = [\theta_4(n), \theta_{4+4m}(n), \theta_{4+8m}(n), \theta_{4+12m}(n) \dots \quad p\theta_d(n)] \qquad (64)$$

and similarly, $\Theta_{|1>}(n)$ and $\Theta_{-|1>}(n)$

The chart on the left side of Figure 5 describes the probability distribution of the zero quantum state, $|0>_c : |\psi_{|0>}(n)|^2$, and above it (in red) the phase distribution $\Theta_{|0>}(n) - \Theta_{-|0>}(n)$. The

chart on the right describes the probability distribution, $|\psi_{|1>}(n)|^2$, and above it (in red) the phase distribution $\Theta_{|1>}(n) - \Theta_{-|1>}(n)$. From the graphs, it seems that the phase difference becomes dominant in both cases near the edge of the system.

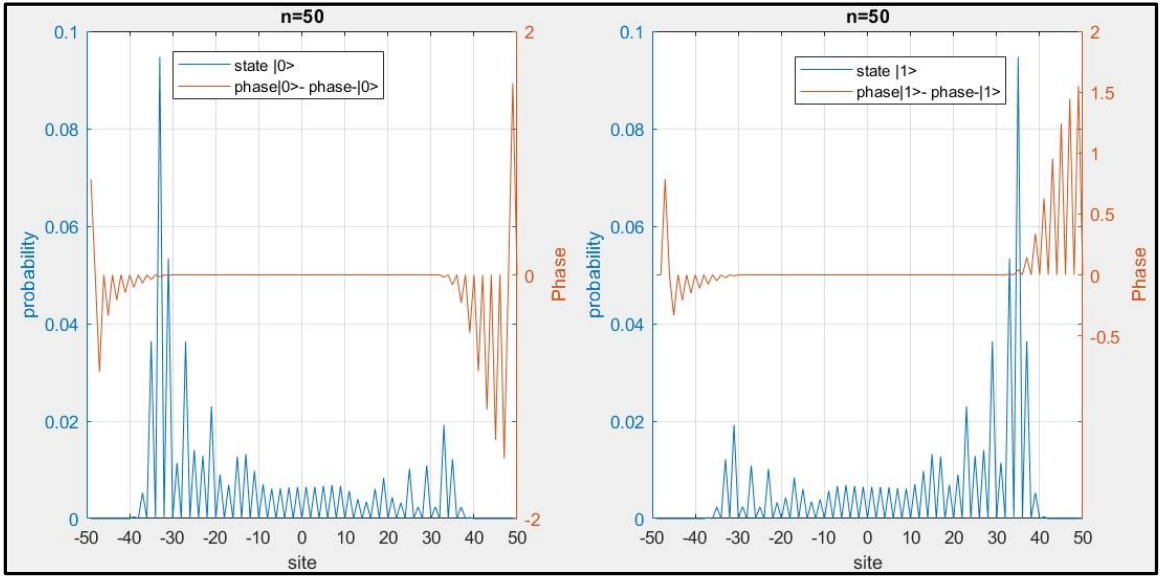

Figure 5. The probability distribution of the quantum states $|0>_c, |1>_c$ after 50 steps and the corresponding phase difference

## Quantum walks on a finite line in space.

It is possible to embed boundary conditions in the model described in Figure 2, such as a trap:

$$Trap = \begin{bmatrix} 0 & 0 & 0 & 0 \\ 0 & 0 & 0 & 0 \\ 0 & 0 & 0 & 0 \\ 0 & 0 & 0 & 0 \end{bmatrix}, \quad (65)$$

or to apply a reflecting point. In the latter case, there are two kinds of reflecting barriers described by switching between the states $|0>_c, |1>_c$ and $-|0>_c, -|1>_c$, presented by

$$R_1 = \frac{1}{\sqrt{2}} \begin{bmatrix} 0 & 1 & 0 & 0 \\ 1 & 0 & 0 & 0 \\ 0 & 0 & 0 & 1 \\ 0 & 0 & 1 & 0 \end{bmatrix}, \quad (66)$$

or switching between the states $|0>_c, -|1>_c$ and between $-|0>_c, |1>_c$, presented by

$$R_2 = \frac{1}{\sqrt{2}} \begin{bmatrix} 0 & 0 & 1 & 0 \\ 0 & 0 & 0 & 1 \\ 1 & 0 & 0 & 0 \\ 0 & 1 & 0 & 0 \end{bmatrix}, \quad (67)$$

The matrices that describe the reflecting boundaries $R_1$ and $R_2$ without the $1/\sqrt{2}$ factor are unitary transformations. This factor will be explained formally; however, the $1/\sqrt{2}$ factor cancels the $\sqrt{2}$ that occurs on each site (The $\sqrt{2}$ was taken out of the transition matrix $A$ to create a stochastic matrix).

The following figure describes a system with two reflecting boundaries at locations -3 and 3:

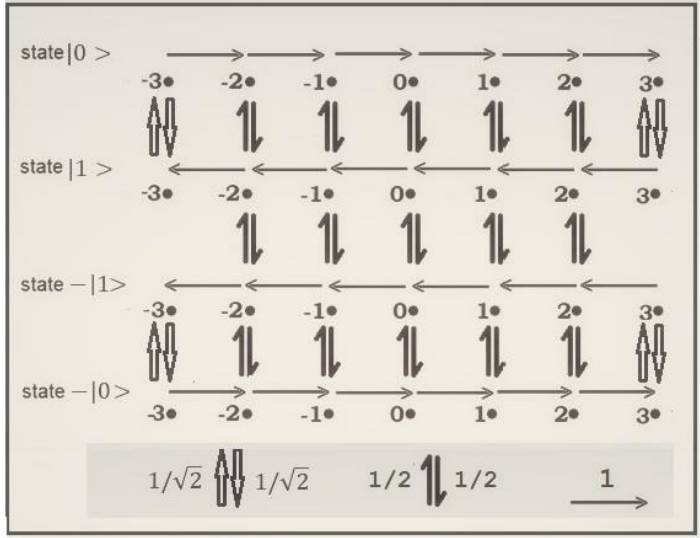

Figure 6. A finite two-step Markovian model. The horizontal direction (x-axis) describes the movements due to the shift operator, and the vertical direction (y-axis) describes the movements due to the transition matrix $A$ and $R_1$ at the boundaries. At the bottom of the figure, there is a description of the transition probability values of each of the arrows.

Note that the x-axis operator is not unitary (almost unitary). In order to make it unitary, it is necessary to add the following connection, i.e., to make the system cyclic with respect to the x-operator as follows:

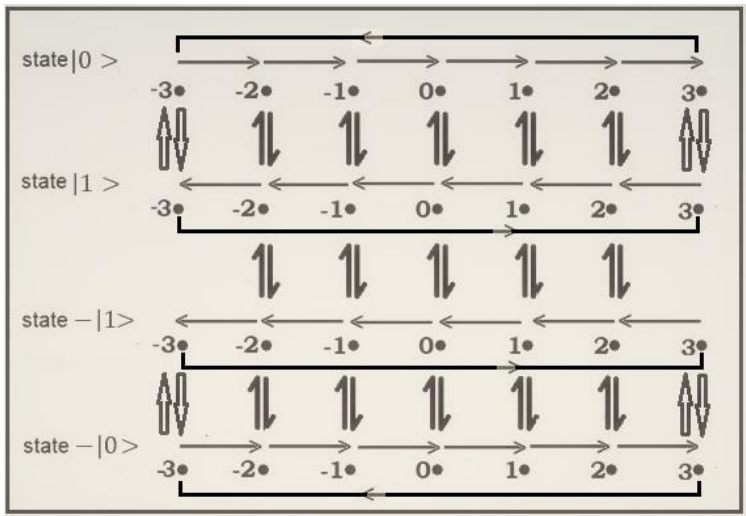

Figure 7. Cyclization of the x-axis operator

So mathematically, the x-axis can be presented by a unitary matrix by adding the entries: $Right(d, 1) = 1, Left(1, d)$. Still, this cyclization is unnecessary, since actually no population propagates from the left side of the system to the other side of the system. This can be explained as follows:

Assume that at step $n$, the one states $|1>_c$ reach site -3, and then the y-axis operates and therefore switches all the quantum one states $|1>_c$ into the zero states $|0>_c$. Then the shift operator operates again, but none of the ones, $|1>_c$, remain on the left boundary. Therefore, nothing propagates to the other side of the system.

The same occurs at the right side, i.e., assume that at step $n$ zero states, $|0>_c$, reach site 3, and then the y-axis operator operates and switches all the $|0>_c$ to $|1>_c$ and no zero states remain to shift to the other side of the system. Therefore, both systems presented in Figures 6 and 7 behave precisely the same when initializing at any location, except when initializing at the boundaries.

Also, note that while the random walker feels a finite system with two partial traps, the quantum walker feels a reflecting barrier, and the unitary property of the system is conserved.

Stated formally: the reflecting boundaries for the Hadamard walk for the case of switching between the states $|0>_c, |1>_c$ can be formulated as follows:

$$r_1 = \begin{bmatrix} 0 & 1 \\ 1 & 0 \end{bmatrix} \tag{68}$$

And for the case of switching between the states of opposite signs $|0>_c, -|1>_c$ or $-|0>_c, |1>_c$, presented by

$$r_2 = -\begin{bmatrix} 0 & 1 \\ 1 & 0 \end{bmatrix} \tag{69}$$

the shift operator can be written as before with:

$$x = Right \otimes Zero + Left \otimes One \tag{70}$$

The Right and the Left matrices were defined previously, and it is also possible (but not necessary) to add the cyclization thus $Right(d, 1) = 1, Left(1, d)$

The y-axis operator can be written as follows:

$$y_{rp} = (I_d - Z) \otimes H + Z \otimes r_1 \tag{71}$$

Where $H$ and $r_1$ are unitary matrices and the $Z$ matrix is a $d \times d$ zero matrix, except for two entries at the beginning and the end of the main diagonal, namely $Z(1,1) = 1; Z(d, d) = 1$.

The dynamics of the Hadamard walk can be written as:

$$u_{rp} = xy_{rp} = (Right \otimes Zero + Left \otimes One)(I_d \otimes H - Z \otimes H + Z \otimes r_1) \quad (72)$$

Returning to the system depicted in Figure 6, the y-axis operator of the system in a two-reflecting-boundaries condition is:

$$Y_{rp} = (I_d - Z) \otimes A + Z \otimes R_1 \quad (73)$$

and the shift operator as before

$$X = Right \otimes Zerostate + Left \otimes OneState \quad (74)$$

Thus the whole dynamics is:

$$U_{rp} = XY_{rp} = (Right \otimes Zerostate + Left \otimes OneState)((I_d - Z) \otimes A + Z \otimes R_1) \quad (75)$$

The equivalence between the Hadamard walk with two reflecting barriers and the Markovian model presented in Figure 6 is proven step-by-step.

Using the previous relations of: $BZerostateB^T = 2Zero$, $BOneStateB^T = 2One$,

$H = \frac{1}{\sqrt{2}} BAB^T$ and the new relations between the reflecting boundaries of the Hadmard walk versus the reflecting boundaries depicted in figure 6 :

$$r_1 = \frac{1}{\sqrt{2}} BR_1 B^T, \quad r_2 = \frac{1}{\sqrt{2}} BR_2 B^T \quad (76)$$

(which can explicitly be proved by multiplication)

And substituting these relations into Eq. (72) yields:

$$u_{rp} = \frac{1}{2\sqrt{2}} (Right \otimes BZerostateB^T + Left \otimes BOnestateB^T)((I_d - Z) \otimes BAB^T + Z \otimes BR_1 B^T) \quad (77)$$

Multiplying both sides by $(I_d \otimes B)$ and using the Kronecker product property for the right side yields

$$u_{rp}(I_d \otimes B) = \frac{1}{2\sqrt{2}} (Right \otimes BZerostateB^T + Left \otimes BOnestateB^T)((I_d - Z) \otimes BAB^T B + Z \otimes BR_1 B^T B) \quad (78)$$

Since $B^T B$ commutes with the transition matrix $A$ and the reflecting boundaries $R_1$, i.e., $[A, B^T B] = 0$, and $[R_1, B^T B] = 0$, then

$$u_{rp}(I_d \otimes B) = \frac{1}{2\sqrt{2}} (Right \otimes BZerostateB^T + Left \otimes BOnestateB^T)(I_d \otimes BB^T BA - Z \otimes BB^T BA + Z \otimes BB^T BR_1 \quad (79)$$

Rearranging Eq. (79) using the Kronecker product property yields:

$$u_{rp}(I_d \otimes B) = \frac{1}{2\sqrt{2}}(Right \otimes BZerostateB^T + \qquad (80)$$

$$Left \otimes BOnestateB^T)(I_d \otimes BB^TB)(I_d \otimes A - Z \otimes A + Z \otimes R_1) =$$

$$\frac{1}{2\sqrt{2}}(Right \otimes BZerostateB^TBB^TB + Left \otimes BOnestateB^TBB^TB)(I_d \otimes A - Z \otimes A + Z \otimes R_1)$$

since $[Zerostate, B^TB] = 0$ and $[OneState, B^TB] = 0$ then:

$$u_{rp}(I_d \otimes B) = \frac{1}{2\sqrt{2}}(Right \otimes BB^TBB^TBZerostate + \qquad (81)$$

$$Left \otimes BB^TBB^TBOnestate)(I_d \otimes A - Z \otimes A + Z \otimes R_1)$$

And using the following property $BB^T = 2I_2$ yields:

$$u_{rp}(I_d \otimes B) = \sqrt{2}(Right \otimes BZerostat + Left \otimes BOnestat)(I_d \otimes A - Z \otimes A + Z \otimes R_1) \qquad (82)$$

And finally, substituting Eq. 75 into Eq. 82 yields:

$$u_{rp}(I_d \otimes B) = \sqrt{2}(I_d \otimes B)U_{rp} \qquad (83)$$

In other words, the system depicted in Figure 6 and Figure 7 described Hadmard walk between two reflecting boundaries.

The following figures describe the probability distribution of a quantum random walk on a finite line of 25 sites after 35 and 65 steps, where each site (except the boundaries) initializes at $(|0> - |1>)/\sqrt{46}$.

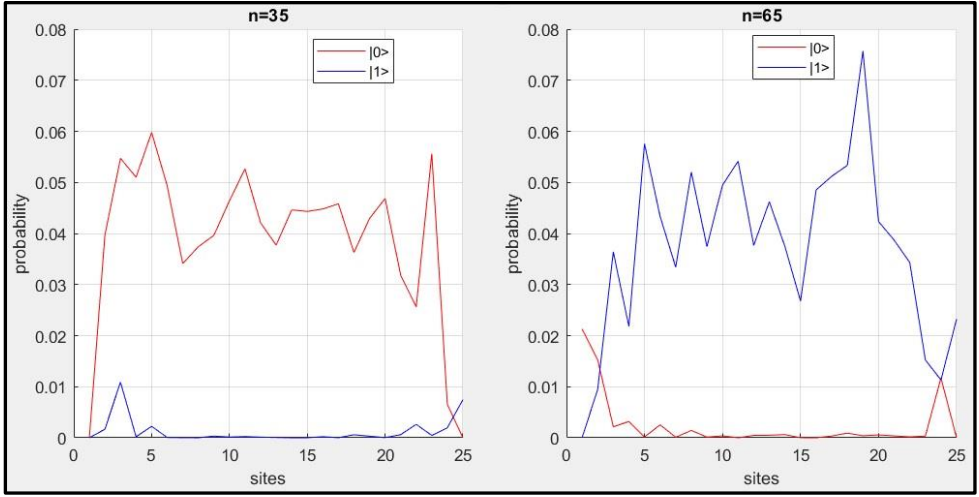

Figure 8. The probability distribution of a quantum random walk on a finite line of 25 sites after 35 and 65 steps, starting at locations 2–24 with the state $(|0> - |1>)/\sqrt{46}$.

Note that the system initializes with equally distributed states |0> and –|1>, then switches after 35 steps mostly to state |0>, and after 65 steps mostly goes back to state |1>.

The system preserves its unitary character; specifically in this case:

$$2^n \|P_{|0>}(n) - P_{-|0>}(n)\|^2 + 2^n \|P_{|1>}(n) - P_{-|1>}(n)\|^2 = 1 \qquad (84)$$

Figure (9) presents the distribution of the entire population, namely the sum of

$P_{|0>}(n) + P_{|1>}(n) + P_{|1>}(n) + P_{-|0>}(n)$ as a function of the kth site. Note that the distribution behaves like a random walk between two partial traps, as expected.

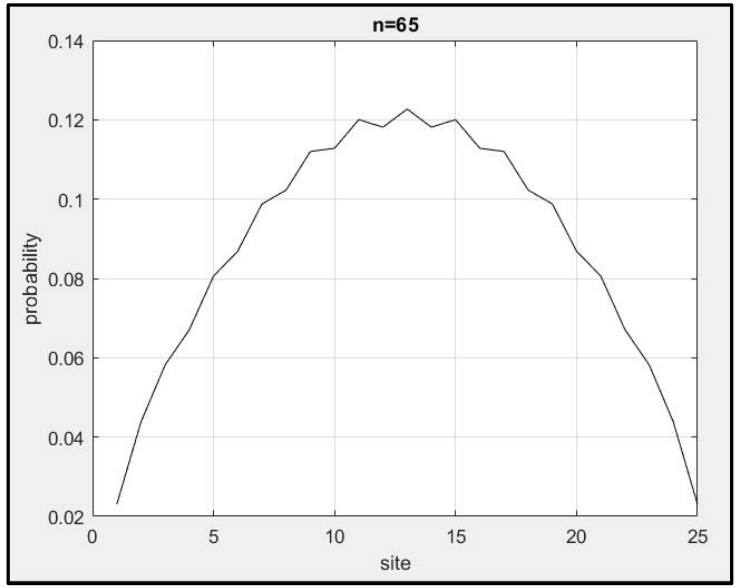

Figure 9. The probability distribution of a random walk on a finite line of 25 sites between two reflecting points after 65 steps, initializing at locations 2–24 with the quantum state of $(|0> - |1>)/\sqrt{46}$

## Summary

Herein, a two-dimensional model that maps both a random and a quantum random walk onto the same system was presented. The system conserves the population distribution for the case of an infinite line, similar to a Markovian process presented by a stochastic matrix. By using a proper transformation, it yields the amplitude distribution of two quantum states |1>, |0>, similar to a unitary operator. The model shows that for a large $n \gg 1$, the amplitudes of each of the four quantum states behave like a Gaussian distribution. The model also enables us to embed different boundary conditions, such as a reflecting point or a trap, and has the advantage of revealing each of the four coin states without interference, thereby acquiring a deeper understanding of the entire process.

# References.

[15] In an explicit way by matrix multiplication, or note $B^T B = I_4 - J_4$ and the reversal matrix commute with any centrosymmetric matrix such as zeroState or oneState ($I_4$ commute with any 4X4 matrices)

[16] From Eq. (37) : $U = Right \otimes ZerostateA + Left \otimes OneStateA$

By using Kronecker product property :

$U(I_d \otimes B^T)(I_d \otimes B) = Right \otimes ZerostateAB^T B + Left \otimes OneStateAB^T B$

$(I_d \otimes B^T)(I_d \otimes B)U = Right \otimes B^T BZerostateA + Left \otimes B^T BOneStateA$

and since $[B^T B, Zerostate] = 0 \ and \ [B^T B, A]$ explicity by multiplication thus:

$U(I_d \otimes B^T)(I_d \otimes B) = Right \otimes B^T BZerostateA + Left \otimes B^T BOneStateA$

Therefore: $U(I_d \otimes B^T)(I_d \otimes B) = (I_d \otimes B^T)(I_d \otimes B)U$